\documentclass[superscriptaddress,aps,pra,reprint,amsmath,amssymb]{revtex4-1}
\usepackage{graphicx}
\usepackage{dcolumn}
\usepackage{bm}
\usepackage{color}
\usepackage{soul}
\usepackage{hyperref}

\begin{document}

\preprint{APS/123-QED}

\title{Microwave Near-Field Quantum Control of Trapped Ions}

\author{U.~Warring}
\altaffiliation{Present address: Albert-Ludwigs-Universit\"at Freiburg,  Physikalisches Institut, Hermann-Herder-Str. 3, 79104 Freiburg, Germany}
\email{ulrich.warring@physik.uni-freiburg.de}
\affiliation{Time and Frequency Division, National Institute of Standards and Technology; 325 Broadway, Boulder, Colorado 80305, USA}

\author{C.~Ospelkaus}
\affiliation{Time and Frequency Division, National Institute of Standards and Technology; 325 Broadway, Boulder, Colorado 80305, USA}\affiliation{QUEST, Leibniz Universit\"at Hannover, Welfengarten 1,
30167 Hannover \\and PTB, Bundesallee 100, 38116 Braunschweig, Germany}

\author{Y.~Colombe}
\author{K.~R.~Brown}
\altaffiliation{Present address: GTRI Georgia Tech, 400 10th Street NW, Atlanta, Georgia 30318, USA}
\author{J.~M.~Amini}
\altaffiliation{Present address: GTRI Georgia Tech, 400 10th Street NW, Atlanta, Georgia 30318, USA}
\affiliation{Time and Frequency Division, National Institute of Standards and Technology; 325 Broadway, Boulder, Colorado 80305, USA}

\author{M.~Carsjens}
\affiliation{QUEST, Leibniz Universit\"at Hannover, Welfengarten 1,
30167 Hannover \\and PTB, Bundesallee 100, 38116 Braunschweig, Germany}

\author{D.~Leibfried}
\email{dil@boulder.nist.gov}
\author{D.~J.~Wineland}
\affiliation{Time and Frequency Division, National Institute of Standards and Technology; 325 Broadway, Boulder, Colorado 80305, USA}

\date{\today}

\begin{abstract}
Microwave near-field quantum control of spin and motional degrees of freedom of $^{25}$Mg$^{+}$ ions can be used to generate two-ion entanglement, as recently demonstrated in Ospelkaus \textit{et al.} [Nature $\mathbf{476}$, 181 (2011)]. Here, we describe additional details of the setup and calibration procedures for these experiments. We discuss the design and characteristics of the surface-electrode trap and the microwave system, and compare experimental measurements of the microwave near-fields with numerical simulations. Additionally, we present a method that utilizes oscillating magnetic-field gradients to detect micromotion induced by the ponderomotive radio-frequency potential in linear traps. Finally, we discuss the present limitations of microwave-driven two-ion entangling gates in our system.
\end{abstract}


                              
\maketitle

\section{Introduction}

Scaling to large numbers of quantum bits (qubits) is important for the development of a future quantum information processor. Several pathways toward this goal have been proposed based on using trapped atomic ions as qubits~\cite{cirac_quantum_1995,wineland_experimental_1998,kielpinski_architecture_2002,blatt_entangled_2008,blatt_quantum_2012,monroe_large_2012}. One particular approach envisions processing, transporting, and sympathetic cooling of ions inside a large array of traps~\cite{wineland_experimental_1998,kielpinski_architecture_2002,blatt_entangled_2008}. Recent experiments with various trap architectures have demonstrated the key ingredients for scalable ion loading, trapping, processing, and transport~\cite{rowe_transport_2002,schulz_sideband_2008, home_complete_2009, hanneke_realization_2009, amini_toward_2010, moehring_design_2011, blakestad_near-ground-state_2011, shaikh_monolithic_2011, gaebler_randomized_2012, walther_controlling_2012, bowler_coherent_2012}.

Current implementations of quantum information processors require a reduction of gate errors to be efficiently scalable~\cite{ladd_quantum_2010}, necessitating stringent qubit control. In most trapped-ion quantum information experiments, control is accomplished via laser-based techniques~\cite{blatt_entangled_2008}, but other approaches are being investigated, e.g., those based on magnetic fields~\cite{mintert_ion-trap_2001,chiaverini_laserless_2008, ospelkaus_trapped-ion_2008}. These methods reduce the required laser overhead and may enable a higher level of control and integration than laser-based control. While the highest entangled-state fidelity is currently obtained with laser-based operations~\cite{benhelm_towards_2008}, magnetic-field based control may eventually provide sufficiently low errors for fault tolerant operations~\cite{knill_resilient_1998,preskill_reliable_1998,steane_overhead_2003,knill_quantum_2005,raussendorf_fault-tolerant_2007}. For single-qubit operations performed with microwave magnetic near fields, errors per computational gate of $2 \times 10^{-5}$ and better have been reported~\cite{brown_single-qubit-gate_2011, allcock_private_2012}. Although the theoretical threshold for fault tolerance depends strongly on the assumptions of the model under study, an operational error of 10$^{-4}$ or below may be a reasonable starting point in a future fault-tolerant architecture. Reaching this level of precision for two-qubit operations requires improvements on the state-of-the-art errors by approximately two orders of magnitude.

As with conventional computers, microfabrication may be advantageous in realizing a future quantum processor, such as those based on surface-electrode ion traps~\cite{seidelin_microfabricated_2006} (for a review see~\cite{hughes_microfabricated_2011}). In addition to reducing the overall volume of a processor for a given number of qubits, a reduction in scale could allow for tighter confinement with trap frequencies scaling approximately as $1/d^2$ if all other parameters are held constant. Here $d$ denotes a characteristic length scale of the trap, often characterized by the smallest ion-electrode distance. Higher trap frequencies can enable shorter durations for transport, separation and recombination, and two-qubit operations mediated by ion motion. In the last two decades, typical trap length scales have been reduced from centimeters to tens of micrometers. This increases the oscillating magnetic near-fields (scaling as $1/d$), originating from oscillating currents running in one or multiple trap electrodes. Fast and precise single-qubit rotations (carrier transitions) have been demonstrated in this way~\cite{ospelkaus_microwave_2011, brown_single-qubit-gate_2011,allcock_private_2012}. In addition, the corresponding field gradients (proportional to $1/d^2$) can become large enough to couple the ions' spin with their motional state (sideband transitions) on experimentally accessible timescales~\cite{ospelkaus_trapped-ion_2008, ospelkaus_microwave_2011}. For trapped ions close to their motional ground states, the ratio of sideband-transition rates to carrier-transition rates driven by oscillating near fields and their corresponding field gradients is on the order of $a_0/d$, where $a_0$ is the harmonic-oscillator ground-state extent~\cite{ospelkaus_trapped-ion_2008}. For typical experimental parameters, $a_0$ is on the order of $10$~nm, and therefore $a_0/d$ ranges between $10^{-3}$ and $10^{-4}$ for the smallest traps today. Further reduction in trap size would increase this ratio, enabling faster spin-motional entanglement. For the same reason, near-field gradients are typically not useful for larger ($d > 100~\mu$m) trap structures.

The main practical limitation to size reduction has been ``anomalous'' motional heating, caused by electric-field noise at the positions of the ions in excess of Johnson noise from the resistances of the electrodes and of their filters~\cite{turchette_heating_2000, daniilidis_fabrication_2011}. Anomalous noise fields have been phenomenologically modeled as originating from potential fluctuations of surface patches or dipoles with typical dimensions much smaller than $d$, which yields a noise spectral density that scales as $1/d^4$~\cite{turchette_heating_2000}. As $d$ is decreased, the resulting heating grows faster than trap frequencies, oscillating near fields, and their corresponding gradients. Consequently, this form of heating appears to prevent scaling to even smaller trap dimensions. However, recent studies have demonstrated a substantial reduction of anomalous heating by cooling the electrodes~\cite{deslauriers_scaling_2006, labaziewicz_suppression_2008, labaziewicz_temperature_2008} and by treatment of the electrode surfaces~\cite{allcock_reduction_2011, hite_100-fold_2012}. Such methods may enable the further reduction in trap size.

In the experiments described here, we investigate building blocks for oscillating magnetic near-field quantum control of trapped ions~\cite{ospelkaus_trapped-ion_2008}; a similar approach is pursued in~\cite{allcock_microfabricated_2012}. To test the basic scheme, we integrated three microwave current-carrying electrodes into a room-temperature surface-electrode trap with $d\simeq 30~\mu$m. Frequencies of the microwave currents are tuned to near resonance with hyperfine transitions in $^{25}$Mg$^{+}$ (nuclear spin $5/2$). The resulting magnetic field at the position of the ions determine the carrier coupling, while the magnetic-field gradient couples the hyperfine levels (spin states) to the motional degrees of freedom~\cite{ospelkaus_trapped-ion_2008}. We previously reported on fast-carrier transitions and generation of two-ion entanglement~\cite{ospelkaus_microwave_2011}; the entanglement was produced with a two-qubit gate operation~\cite{milburn_ion_2000,soerensen_quantum_1999,solano_deterministic_1999} suitable for scalable quantum information processing. In this report, we give a more detailed account of our experimental setup and the procedures for calibrating and optimizing the operation of the system.

In Sec.~\ref{sec:trap}, we discuss design, fabrication, and performance of the surface-electrode trap and describe the laser beams for loading, state preparation, and detection. Section~\ref{sec:MWsetup} describes the microwave system that generates and shapes the signals used for coherent single-qubit and two-qubit operations. In Sec.~\ref{sec:procedure}, we give an experimental sequence for calibrating and studying entangling-gate operations. The experimental procedure to adjust the microwave field at the ion's position for driving sideband transitions, a prerequisite for entangling gates~\cite{milburn_ion_2000,soerensen_quantum_1999,solano_deterministic_1999},  is described in Sec.~\ref{sec:OscB}. Section~\ref{sec:MM} details a technique for rf micromotion detection by observing transitions in microwave magnetic-field gradients. In Sec.~\ref{sec:radialMode}, we describe a procedure to optimize the orientation of the radial-motional modes relative to the microwave field gradient components and a static homogeneous quantization field. Finally, in Sec.~\ref{sec:limits} we discuss the factors currently limiting two-qubit gate fidelity~\cite{ospelkaus_microwave_2011}.
\section{Surface-Electrode Trap and Experimental Setup}
\label{sec:trap}
\begin{figure}
  \centering
  \includegraphics[width=\columnwidth]{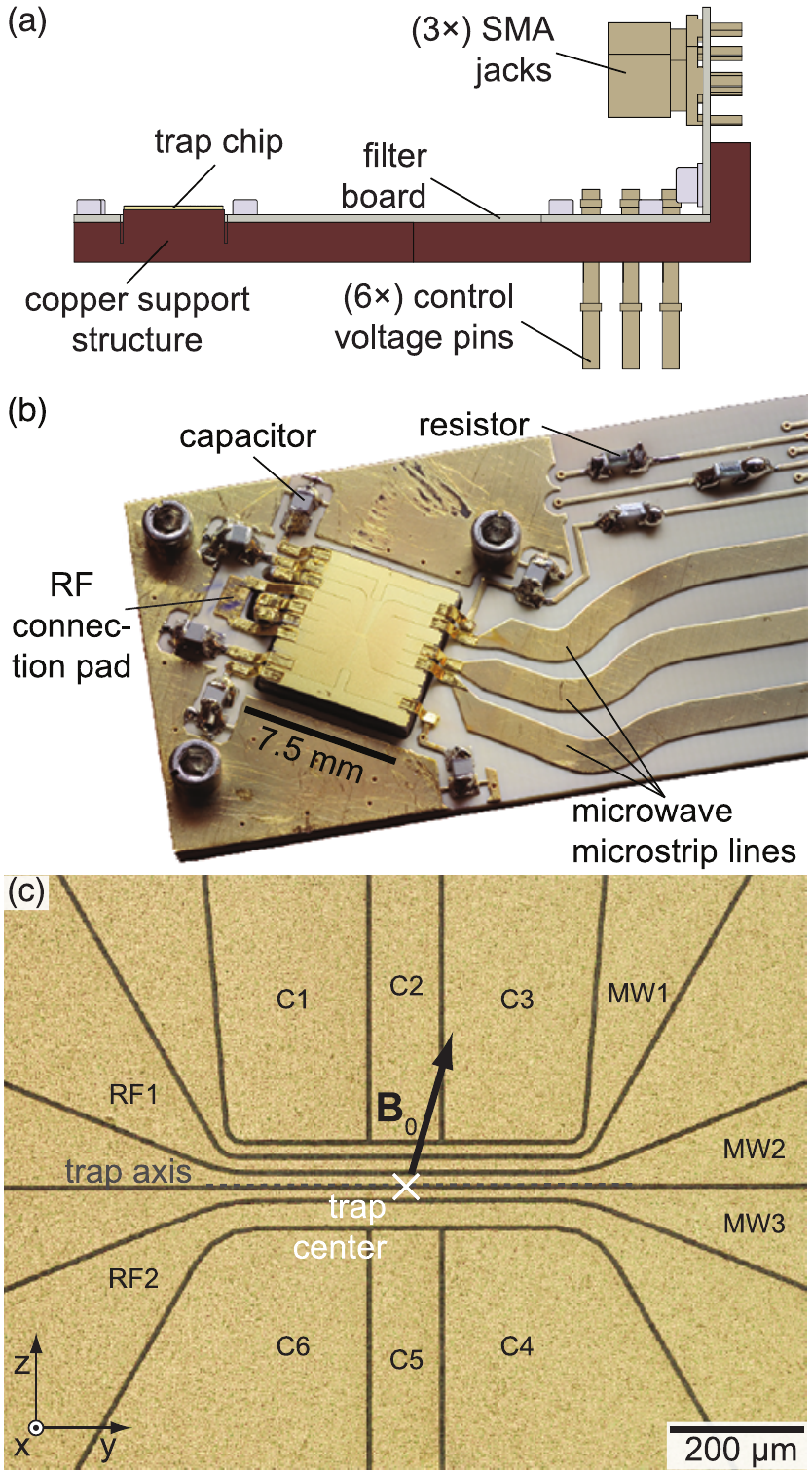}
  \caption{(Color online) (a) Cross-section view of the filter board and trap chip assembly. (b) Photograph of filter board and mounted surface-electrode trap. (c) Micrograph of the central region of the surface-electrode trap. The electrodes are labeled. The direction of the external quantization field $|\bold{B_{\text{0}}}| \simeq 21.28$~mT, located in the $y$-$z$ plane, is shown. The $^{25}$Mg$^{+}$ ions are trapped at a distance $d \simeq 30~\mu$m above the surface, at the position labeled trap center.}
\label{fig:chip}
\end{figure}

The surface-electrode trap is designed as an asymmetric ``five-wire'' trap~\cite{chiaverini_surface-electrode_2005, wesenberg_electrostatics_2008}. It consists of two radio-frequency (rf), six control, and three microwave electrodes; see Fig.~\ref{fig:chip} for a view of the trap assembly (Fig.~\ref{fig:chip}a, b) and the central region of the trap electrodes (Fig.~\ref{fig:chip}c).

The trap chip is fabricated in the NIST Boulder cleanroom. We choose aluminum nitride (AlN) as a substrate for its high thermal conductivity of approximately $170$~W/(m$\cdot$K), which aids dissipation of heat generated by the microwave currents in the trap electrodes. Adhesion (Ti, 10~nm) and seed (Cu, 60~nm) layers are deposited on a $75$~mm diameter AlN wafer by electron beam evaporation. A 14 $\mu$m thick layer of photoresist (Shipley SPR220-7~\footnote{Any mention of commercial products is for information only; it does not imply recommendation or endorsement by NIST.}) is spin-coated onto the wafer, exposed through a patterned chromium mask by contact lithography, and developed. The photoresist pattern defines $4.5~\mu$m wide gaps between electrodes in the central region of the trap; the gap width is increased to $25~\mu$m at distances $\simeq 2$~mm away from the trap center. Gold is then electroplated, in a 140~mL sulfite gold solution (Transene TSG-250~\cite{Note1}), up to a $10.5~\mu$m thickness onto individual approximately $22.5~\text{mm}\,\times\, 22.5$~mm dies. The use of a small electroplating volume allows us to replace the solution for each electroplating process, thus reducing the possibility of contamination. After removing the photoresist with acetone, the seed and adhesion layers are wet etched, and a total of nine (approximately $7.5~\text{mm}\,\times\,7.5$~mm) trap chips are obtained by dicing. To provide good thermal contact, the chip is bonded with a thin layer of silver-filled vacuum compatible epoxy (EPO-TEK$^{\text{\textregistered}}$~H21D~\cite{Note1}) to a solid copper support (Fig.~\ref{fig:chip}a). This support structure also holds a printed circuit board for in-vacuum filtering of the control potentials. We use low-pass RC filters with $R = 1$\,k$\Omega$ (ANAREN, part no.: H2B15081001F2LO~\cite{Note1}) and $C=820$\,pF (NOVACAP, part no.: 0504N821J101P~\cite{Note1}). In addition, the printed circuit board is used for connecting both rf electrodes to a resonant quarter-wave step-up transformer ($Q_{\text{res}} \simeq 350$, when loaded with the trap)~\cite{jefferts_coaxial-resonator-driven_1995}. Connections between all trap chip electrodes and the filter board are made with gap-welded gold ribbons with a $50~\mu$m$\,\times\,500~\mu$m cross section and length $\leq 2$~mm. Each microwave electrode is connected to a microstrip line on the filter board, which is soldered to a SMA jack on the input end. The other end of each microwave electrode is shorted to ground at the edge of the chip via gap-welded gold ribbons. In-vacuum coaxial cables connect the vacuum feedthroughs with the SMA connectors. All of the assembly steps are performed in a cleanroom environment to minimize the presence of contamination, e.g., dust particles on the surface of the trap chip, which can generate unwanted conduction paths and stray electric fields, possibly enhanced when exposed to UV laser beams.

An rf peak voltage $V_{\text{RF}} \simeq 15$~V to $60$~V at $\Omega_{\text{RF}} \simeq 2\pi \times 71.6$~MHz is applied to both rf electrodes, providing the $x$-$z$ radial confinement of the ions at a distance $d \simeq 30~\mu$m above the surface. The rf frequency is stabilized to the center of the quarter-wave resonance of the resonator with a feedback loop to the frequency modulation input port of the rf generator. The six control electrodes (C1 to C6) are biased between $- 10$~V and $+10$~V to provide confinement along the trap axis. The potentials are applied by six digital-to-analog converters (DACs; for details see Sec.~\ref{sec:MWsetup}). Single-ion motional frequencies range between $f_{\text{axial}} = 0.5$~MHz and $2.0$~MHz in the axial direction, while two non-degenerate frequencies between $f_{\text{radial}} = 3$~MHz and $12$~MHz aligned along adjustable directions (Sec.~\ref{sec:radialMode}) in the radial $x$-$z$ plane are used. 

We calculate the trapping potential $\phi_{\text{trap}}$ in the radial and axial directions by means of the gapless plane approximation~\cite{wesenberg_electrostatics_2008}; $\phi_{\text{trap}}$ is generated by combining the rf pseudopotential and the control potential.  
\begin{figure}
  \centering
  \includegraphics[width=\columnwidth]{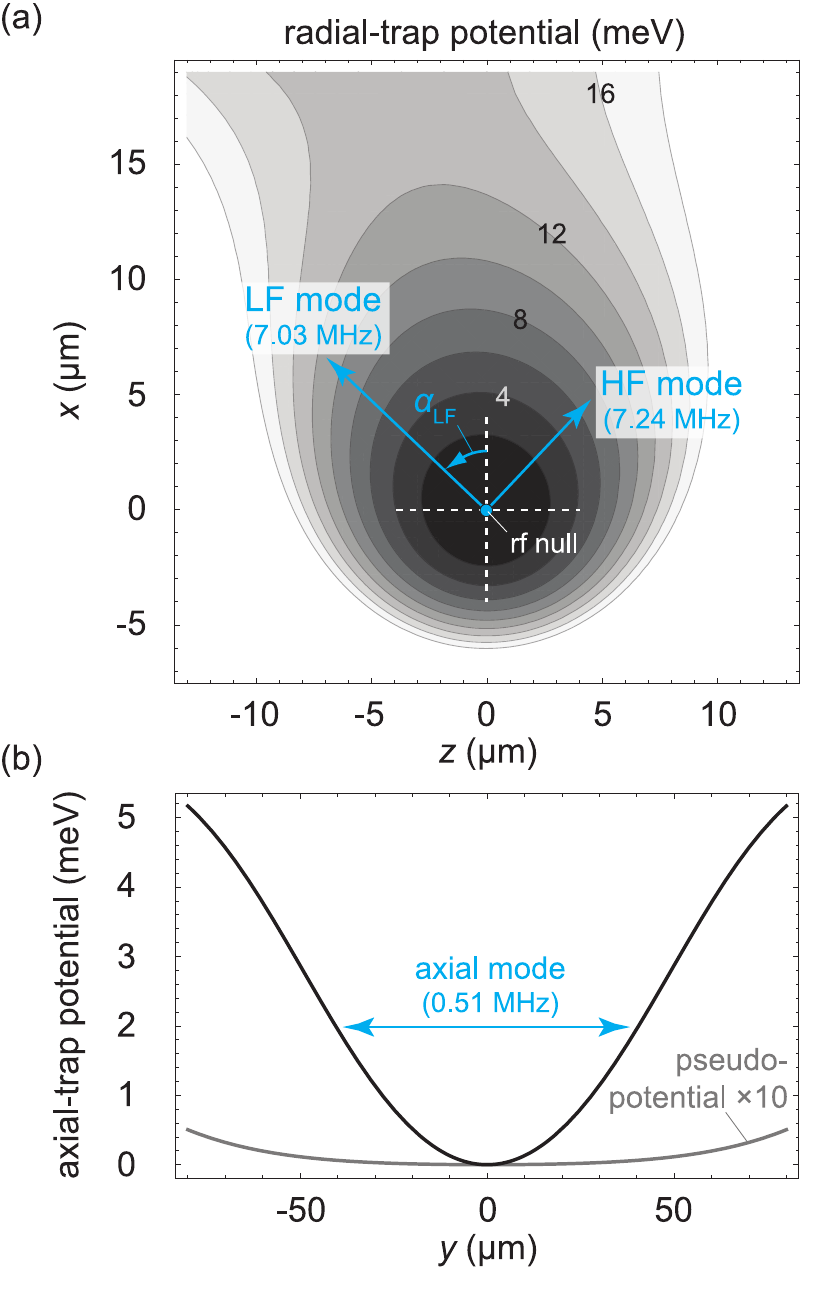}
  \caption{(Color online) Calculated trap potential $\phi_{\text{trap}}$ (in units of meV) as a function of distance from the trap center (rf null and axial minimum), which is located $30~\mu$m above the surface. Applied potentials are given in the text. (a) Potential in radial plane at $y = 0$. The dot marks the trap center. The vectors indicate the calculated motional mode orientation. Here $\alpha_{\text{LF}}$ denotes the angle between the LF-radial mode vector and the $x$ axis. (b) Potential in the axial direction at $x = z = 0$. The contribution from the axial rf pseudopotential (see text) is scaled by a factor of ten for visibility.} 
\label{fig:trappot}
\end{figure}
With $V_{\text{RF}} = 35$~V and $V_{\text{C1-C6}} = \{-0.801, 0.641, -0.801, 0.750, -0.384, 0.750\}$~V, Fig.~\ref{fig:trappot}a shows the calculated $\phi_{\text{trap}}(x,z)$ in the radial plane around the trap center at $x= y= z=0$, while Fig.~\ref{fig:trappot}b illustrates $\phi_{\text{trap}}(y)$ in axial direction. The overall trap depth, limited by a saddle point in the $x$-$z$ plane (Fig.~\ref{fig:trappot}a), is $\simeq 13$~meV for this choice of potentials. We denote the two radial center-of-mass modes by their relative frequencies: low-frequency (LF) and high-frequency (HF) mode; their orientations are indicated by arrows in Fig.~\ref{fig:trappot}a. The orientation of these modes is determined by the control potentials while the rf pseudopotential is cylindrically-symmetric in a small volume around the trap center. Due to the tapered geometry of the rf electrodes, we find a weak axial rf pseudopotential, with a curvature corresponding to a motional frequency $f_{\text{axial}} \simeq 20$~kHz. The axial confinement, however, is dominated by the applied control potentials and its frequency is simulated to be $f_{\text{axial}} \simeq 0.51$~MHz. Additionally, we notice the presence of significant stray potentials, which must be taken into account in the simulations to reproduce the observed motional-mode directions and frequencies (Sec.~\ref{sec:radialMode}).
 
We employ a set of ``shim'' fields, designed through simulations to produce orthogonal displacements of the ion(s) along $x$, $y$, and $z$ while approximately preserving the control-field curvatures, for micromotion nulling (Sec.~\ref{sec:MM}) and to obtain oscillating magnetic-field maps (Sec.~\ref{sec:OscB}). Due to the limited DAC resolution (Sec.~\ref{sec:MWsetup}), the ion position can be controlled to a precision of $\{\Delta x, \Delta y, \Delta z\} \simeq \{16, 37, 5\}$~nm for typical trap frequencies $f_{\text{radial}}\simeq 7.0$~MHz and $f_{\text{axial}} \simeq 1.5$~MHz.

\begin{figure}
  \centering
  \includegraphics[width=\columnwidth]{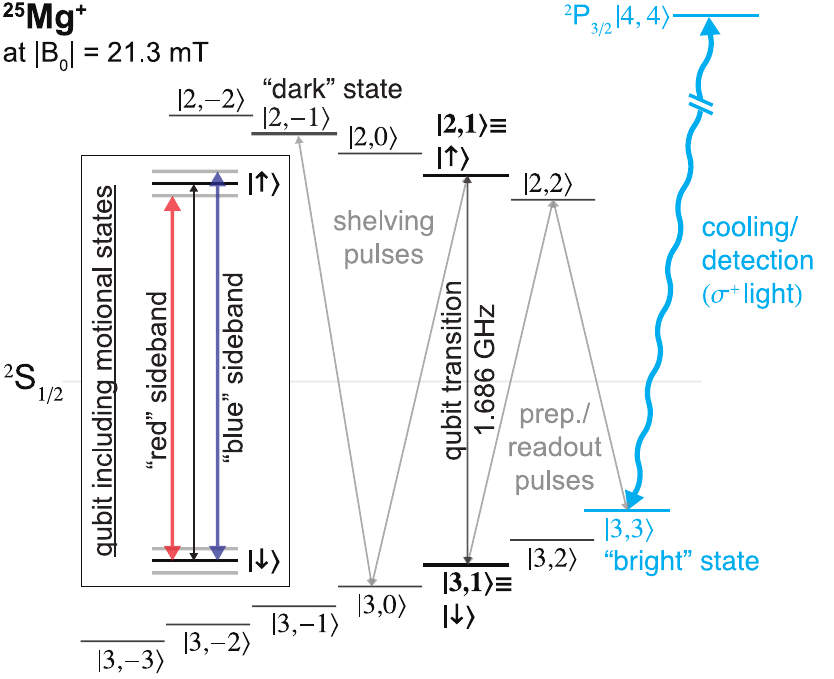}
  \caption{(Color online) Ground state hyperfine structure of $^{25}$Mg$^{+}$ (nuclear spin $5/2$) at $|\bold{B_{\text{0}}}| \simeq 21.3$~mT. The microwave driven hyperfine transitions used to prepare, detect, and control the qubit states are depicted as arrows. The cooling and detection laser beam is indicated as a (blue) wavy arrow. (Inset) Illustration of the qubit states (bold lines) and their radial motional states (gray). The ``red'' and ``blue'' sideband transitions as well as the qubit transition are indicated with arrows.} 
\label{fig:level}
\end{figure}

A static magnetic field of magnitude $|\bold{B_{\text{0}}}| \simeq 21.3$~mT is aligned parallel to the trap surface at an angle of $-15^{\circ}$ with respect to the $z$ axis (Fig.~\ref{fig:chip}c) and provides the ions' internal-state quantization axis. At this field strength, the ground state $^2$S$_{1/2}$ $\left|F = 3, m_{F} = 1\right> \equiv \left|\downarrow\right>$ to $\left|F = 2,m_{F} = 1\right>\equiv \left|\uparrow\right>$ hyperfine transition frequency $f_{\text{0}} \simeq 1.686$~GHz is first-order insensitive to magnetic field changes, while the quadratic frequency deviation is $\simeq 233$~kHz/(mT)$^2$; we use the $\left|\uparrow\right>$ and $\left|\downarrow\right>$ states as the qubit. Here, $F$ is the total angular momentum and $m_{F}$ is the projection of the angular momentum along the magnetic field axis. Figure~\ref{fig:level} illustrates the ground-state hyperfine structure, highlighting the qubit states.
We track $|\bold{B_{\text{0}}}|$-field drifts every 20 to 30 minutes with a Ramsey experiment on the $\left|3,3\right>$ to $\left|2,2\right>$ transition (excited with microwave fields), which has a magnetic-field sensitivity of $\simeq 19.7$~MHz/mT and adjust a computer-controlled correction coil current accordingly. We observe day-to-day $|\bold{B_{\text{0}}}|$-field variations of less than $0.02$~mT.

Two overlapping $\sigma^{+}$-polarized laser beams for Doppler cooling on the $^2$S$_{1/2} \left|3,3\right> \rightarrow ^2$P$_{3/2}\left|4,4\right> $ cycling transition propagate parallel to the magnetic field. A first beam (BDD), approximately $-240$~MHz detuned from this transition, performs initial Doppler cooling and optical pumping. A second beam (BD) is red detuned from this transition by approximately half the natural line width ($\Gamma/2 \simeq 2\pi \times 20$~MHz) and is used for final Doppler cooling and state preparation into the $\left|3,3\right>$ level of the $^2$S$_{1/2}$ ground state~\footnote{Note that here we choose the $\left|3,3\right>$ state for preparation instead of the $\left|3,-3\right>$ state used in~\cite{ospelkaus_microwave_2011}.}. State preparation and cooling takes $1$~ms; during the final $50~\mu$s, only the BD beam is applied. For state detection, the BD beam is tuned to resonance for approximately $200~\mu$s, discriminating the $\left|3,3\right>$ state from the other hyperfine ground states through resonance fluorescence. The fluorescence photons are detected by a photon-multiplier tube (PMT) detector. In most experiments, the $\left|3,3\right>$ population is transferred via two microwave preparation pulses to $\left|\downarrow\right>$ to initialize the qubit. To detect the $\left|\downarrow\right>$ state, two pulses are used to transfer the $\left|\downarrow\right>$ population to the $\left|3,3\right>$ (bright) state, while the $\left|\uparrow\right>$ population is ``shelved'' into the $\left|2,-1\right>$ (dark) state (Fig.~\ref{fig:level}). During the detection period, the PMT registers $\simeq 11$ counts for the bright state and $\simeq 0.2$ counts for the dark state on average per ion. Stray light accounts for $\simeq 0.1$ counts on average. The BDD and BD beams are both focused into the vacuum chamber with a 10-cm focal length lens, and their powers are adjusted to be $\simeq 1~\mu$W and $\simeq 30~\mu$W, respectively; both have waists~\footnote{denotes the $1/e^2$ half width} of $\simeq 15~\mu$m at the position of the ions. To create ions, a third continuous-wave (CW) laser beam (denoted PI: $\lambda \simeq 285$~nm, power $\simeq 3$~mW, and beam waist $\simeq 15~\mu$m) is directed antiparallel to the cooling beams and is used to photoionize neutral Mg atoms inside the trapping region. An oven tube, holding enriched $^{25}$Mg, is mounted approximately $2$~cm away from the trap center. After allowing the oven to heat up for 30~s, the PI is applied for a few seconds until ion(s) are loaded. The oven current is adjusted to a loading rate of less than one ion per second, so the PI beam can be turned off manually after the desired number of ions are detected on a CCD camera. At a background pressure of $2 \times 10^{-9}$~Pa, the storage period of a single ion is observed to be up to a few hours, with Doppler cooling applied periodically every one or two~milliseconds. Typical trapping lifetimes for two ions are on the order of 30~min.

\section{Microwave System}
\label{sec:MWsetup}
\begin{figure}
  \centering
  \includegraphics[width=\columnwidth]{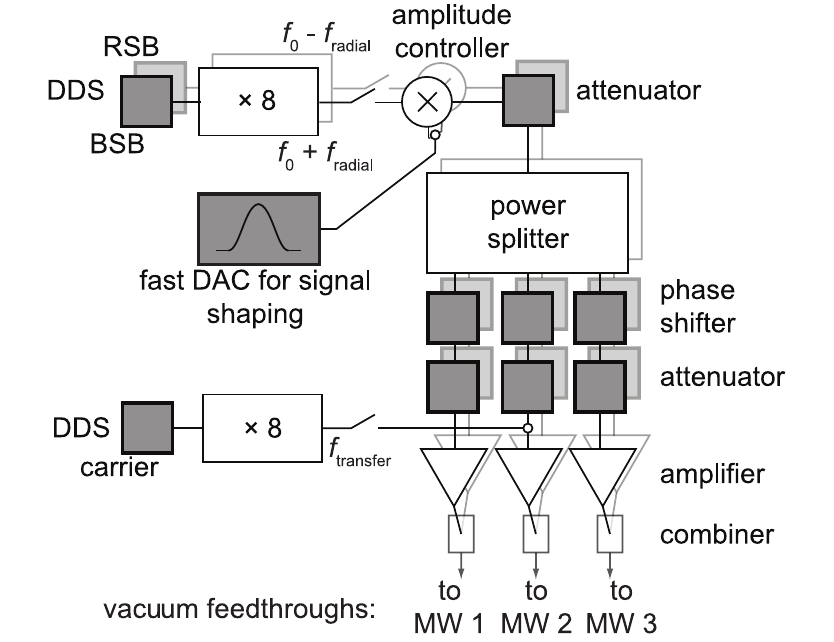}
  \caption{Schematic diagram of the microwave system used to control the microwave fields applied to the ion(s). It enables individual control of phases and amplitudes of the currents in all three microwave electrodes for two frequencies $f_{\text{s}} = f_{0} \pm f_{\text{radial}}$ (BSB and RSB) around $1.69$~GHz, within a range of  about $200$~MHz. It also enables a signal between $1.2$~GHz and $2.3$~GHz applied to MW2 for hyperfine state control and single-qubit operations. The gray components are computer controlled.} 
\label{fig:mwsetup}
\end{figure}

To implement microwave control of the ion(s) spin and motional states, we built a system assembled mostly from commercially available discrete microwave components (Fig.~\ref{fig:mwsetup}). 
For hyperfine state manipulation and single-qubit operations, magnetic fields from microwave currents in electrode MW2 drive transitions at frequencies between $1.2$~GHz to $2.3$~GHz, spanning all hyperfine transitions in the  $^2$S$_{1/2}$ manifold. As noted above, we use a series of two transfer pulses (carrier $\pi$-pulses) to transfer population from $\left|3,3\right>$ to $\left|3,1\right>$ via $\left|2,2\right>$, to initialize the qubit state. We have demonstrated $\pi$-times for qubit spin flips as short as $\simeq 20$~ns~\cite{ospelkaus_microwave_2011}; however, here we adjust the power of the microwave signals to levels corresponding to $\pi$-times on the order of a few microseconds, limited by the timing resolution of our data acquisition system (see below). 

To drive ``blue'' (BSB) and ``red'' (RSB) sidebands of the qubit transition (Fig.~\ref{fig:level}), we apply currents simultaneously to MW1, MW2, and MW3 at $f_{\text{s}} = f_{0} + f_{\text{radial}}$ or $f_{\text{s}} = f_{0} - f_{\text{radial}}$. The amplitudes and phases of the three signals are adjusted to minimize the oscillating magnetic field at the position of the ion(s), while maximizing the field gradient (Sec.~\ref{sec:OscB}). This control is enabled by the six phase shifters and six attenuators.

The microwave setup is integrated with a data acquisition system, controlled by a field-programmable gate array (FPGA) with a timing resolution of $16$~ns~\cite{langer_high_2006}. The entire system comprises 16 digital (TTL) output channels, 24 DAC channels, eight direct-digital synthesizer (DDS) modules, and two digital (TTL) inputs for registering PMT counts. Each DDS can generate signals at frequencies up to $400$~MHz with a frequency resolution of $0.233$~Hz and phase resolution of $0.022^{\circ}$. The DACs have an update rate of $500$~kHz and can be set between $-10$~V and $+10$~V with a resolution of $0.305$~mV. We use six of these channels to apply potentials to the trap control electrodes and 14 channels to control microwave components. In addition, a fast, two-channel DAC is used to generate arbitrary waveforms with a $50$~MHz update rate, a voltage range from $-10$~V to $10$~V, and a resolution of $0.305$~mV. This DAC is programmed via USB and triggered by the data acquisition FPGA.

We use three DDS modules as sources for three frequency octupling modules, each consisting of three doubling stages in series with amplifiers and filters. Two of these generate the RSB and BSB signals, which pass through individual pulse shaping and control stages (phase and amplitude control). The pulse envelopes for the BSB and RSB signals are produced with a multiplier (Analog Devices ADL5391~\cite{Note1}) used to shape pulse amplitudes and controlled by one channel of the fast DAC. In addition, a voltage-controlled attenuator adjusts the overall power of a pulse. After this pulse-shaping stage the signals are split and directed to three individual voltage-controlled phase shifters ($\simeq 12^{\circ}$/V) and attenuators ($\simeq 1.6$~dB/V), to control the power and phase of the signals delivered to the individual microwave electrodes. Before the RSB and BSB signals are combined onto the inputs of each microwave electrode, they are amplified by high-power amplifiers (Mini-Circuits ZHL-30W-252-S+~\cite{Note1}). Following three high-power combiners, three coaxial cables connect the microwave setup with the vacuum feedthroughs. The third octupling module generates frequencies to drive the hyperfine carrier transitions, and its signal is combined with one of the sideband lines attached to electrode MW2. Its power is adjusted with fixed attenuators to yield carrier $\pi$-pulse durations of a few microseconds.  

On a daily basis, we calibrate the frequencies and $\pi$-pulse durations for the transfer and single-qubit pulses. From day to day, we observe fractional variations of $\leq 10^{-3}$ for the pulse durations. Due to our quantization field calibrations and adjustments, the hyperfine-transition frequency values stay constant within a statistical uncertainty of approximately $0.3$~kHz.
\section{Experimental procedures and calibration sequences}
\label{sec:procedure}
\begin{figure}
  \centering
  \includegraphics[width=\columnwidth]{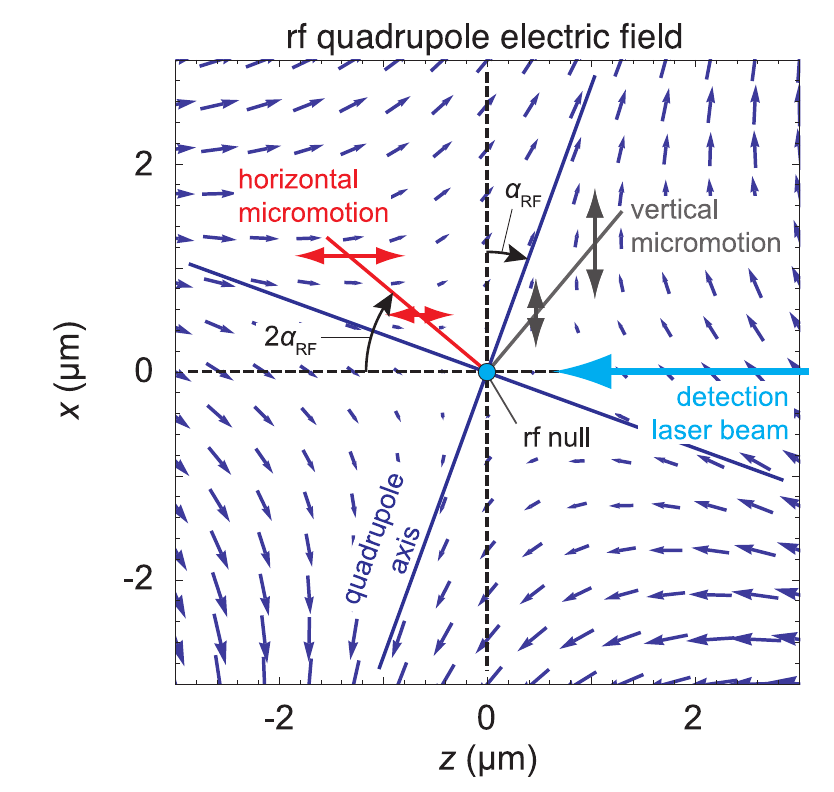}
  \caption{(Color online) Calculated trap rf quadrupole electric field in the radial plane. The quadrupole axes are rotated by $\alpha_{\text{RF}} \simeq -20^{\circ}$. The arrows indicate the horizontal (red) and the vertical (gray) micromotion component. Only the horizontal component can be detected by the detection laser beam. The red and gray lines indicate the two shim directions used for the compensation of horizontal and vertical micromotion.} 
\label{fig:rffield}
\end{figure}

We use a single ion to calibrate experiments and set parameters for the two-qubit gate. After an ion is loaded into the trap, micromotion along the projection of the detection laser beam in the radial plane ($z$ direction) is minimized by measuring the variation in laser fluorescence from the BD beam~\cite{berkeland_minimization_1998} as a function of the ``horizontal''-shim field (Fig.~\ref{fig:rffield}). Since the trap rf quadrupole is tilted in the radial plane by $\alpha_{\text{RF}} \simeq -20^{\circ}$ from the $x$ axis (Fig.~\ref{fig:rffield}), the horizontal-shim field is chosen to push along a vector rotated by $2 \alpha_{\text{RF}}$ from the $z$ axis. A push along this axis will result in a maximal variation of the horizontal micromotion amplitude.
We verify this compensation field with a nulling scheme using microwave magnetic-field gradients (described in Sec.~\ref{sec:MM}). This technique can also be used to minimize the micromotion in the vertical ($y$) direction. We find that the trapping fields are perturbed by stray fields, with corresponding potential $ \phi_{\text{stray}}$. The exact origin of these fields is unknown, but they are consistent with a contact potential patch, and can be compensated for as described in Appendix~\ref{sec:trapPerform}.

Following micromotion nulling, we calibrate the microwave magnetic fields used for driving the radial sidebands of the qubit transition (Sec.~\ref{sec:OscB}). From these experiments, we extract the orientation and the strength of the microwave magnetic-field gradient. As a last calibration step to set up the two-qubit gate experiment, we align the radial-mode vectors relative to the microwave magnetic field by rotating the modes with an additional control field (Sec.~\ref{sec:radialMode}). 

\subsection{Controlling and Mapping the Microwave Near-Field}
\label{sec:OscB}
\begin{figure}
  \centering
  \includegraphics[width=\columnwidth]{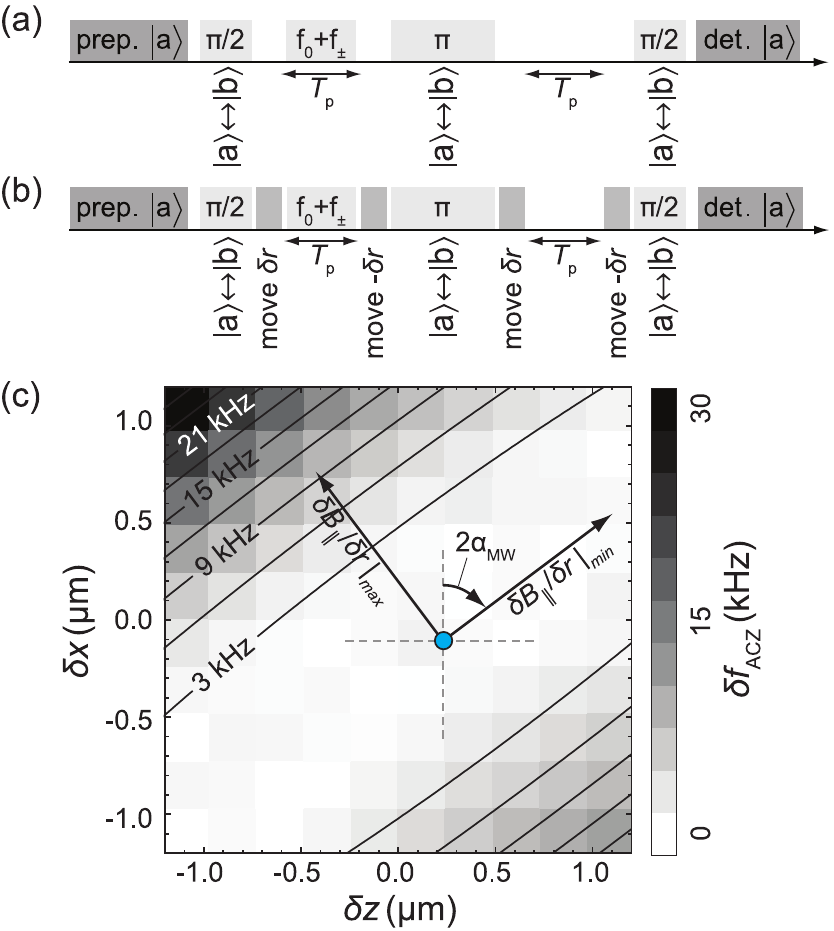}
  \caption{
  Realization of optimal microwave-field configurations for driving motional sideband transitions. 
  (a) Pulse sequence to detect the presence of residual oscillating magnetic fields at the ion(s) position through phase accumulation from the corresponding ac Zeeman shift, while the ion is in a superposition state of $\left|a\right>$ and $\left|b\right>$. The spin echo $\pi$-pulse makes the result insensitive to magnetic-field fluctuations slower than the time scale of an individual experiment (typically $\simeq 500~\mu$s). We observe minimal phase pickup when the oscillating magnetic field is minimized at the position of the ion. (b) Extension to measure the ac Zeeman shift as a function of displacement $\delta r$ (see text). (c) Experimental data mapping out the ac Zeeman shift in the radial $x$-$z$ plane by use of the superposition state of $\left|a\right>=\left|2,0\right>$ and $\left|b\right>=\left|3,1\right>$. The fitted isolines are obtained from the quadrupole oscillating magnetic field component $B_\|$ along $\bold{B}_0$ through its resulting ac Zeeman shift $\delta f_{\text{ACZ}}$, which is proportional to $B_\|^2$. The angle $\alpha_{\text{MW}}$ is the rotation angle of the microwave quadrupole field, but the directions of minimum and maximum values of $|\delta B_\|/\delta r|$ are along angles $2 \alpha_{\text{MW}}$ and $2 \alpha_{\text{MW}} + 90^{\circ}$ (Fig.~\ref{fig:mwfields}a).}
\label{fig:gradmap}
\end{figure}

To implement motional sideband transitions at frequencies $f_{\text{s}}$, a gradient of the microwave magnetic field is necessary. However, any oscillating field at the ion(s) position will cause off-resonant carrier transitions and ac Zeeman shifts~\cite{ospelkaus_trapped-ion_2008}, both of which will inhibit precise control. Thus, the magnetic field amplitude should be suppressed as much as possible at the position of the ion. We realize this experimentally by adjusting the phases and amplitudes of the three microwave currents to minimize ac Zeeman shifts detected by the ion. The experimental sequence is shown in Fig.~\ref{fig:gradmap}a.

The nulling procedure relies on the measurement of ac Zeeman shifts on superpositions of selected hyperfine states $\left|a\right>$ and $\left|b\right>$. In a coordinate system with unit vectors $\bold{e}_{\tilde{z}} = \bold{B_{\text{0}}} / | \bold{B_{\text{0}}} |$, $\bold{e}_{\tilde{x}} = \bold{e}_{x}$ and $\bold{e}_{\tilde{y}} = \bold{e}_{\tilde{z}} \times \bold{e}_{\tilde{x}}$, the magnetic-field components oscillating at $f_\mathrm{s}$ can be written as:
\begin{align}
\label{eq:oscbfieldwrB}
\nonumber
  \bold{\tilde{B}_{\text{MW}}} = 
 &\frac{1}{2} e^{2\pi i f_s t}
	\left(
		\begin{array}{ccc}
			\cos\beta & -\sin\beta & 0\\
			\sin\beta &  \cos\beta & 0\\
			0         &           & 1
		\end{array}
	\right)
\,
  \left(
	\begin{array}{c}
		B_\perp e^{i\gamma}\cos\epsilon \\
	  	B_\perp i e^{i\gamma}\sin\epsilon \\
	  	B_\| 
	\end{array}
	\right) \\
	&+ \mathrm{c.\,c.}
\end{align}
where $B_\|$ and $B_\perp$ are the strengths of the oscillating magnetic-field components parallel and perpendicular to the quantization field $\bold{B_{\text{0}}}$. Here $\beta$ is the rotation angle of the perpendicular magnetic-field component polarization ellipse with the $\bar{x}$ axis and $\gamma$ is a phase. For $\epsilon=0$, the oscillating magnetic field in the plane perpendicular to $\bold{B_{\text{0}}}$ is linearly polarized. For $\epsilon=\pm\pi/4$, the perpendicular field component has pure circular polarization. The $B_\|$ field couples levels with $\Delta m_F=0$, whereas the $B_\perp$ field drives transitions with $\Delta m_F=\pm 1$. These fields induce an ac Zeeman shift $\delta f_{\text{ACZ}} = c_\perp B_\perp^2 + c_\| B_\|^2$ on the $\left| a \right> \leftrightarrow \left| b \right>$ transition, where the coefficients may depend on $f_s$ and~$\epsilon$. Note that we implicitly include the Bloch-Siegert shifts (energy shift resulting from the counter-rotating part of the field) in the calculation of the coefficients in $\delta f_{\text{ACZ}}$.

First consider $\left|a\right>\equiv\left|2,0\right>$ and $\left|b\right>\equiv\left|3,0\right>$. A $B_\perp$ component applied near $f_0$ is red detuned from the $\left|2,0\right>\rightarrow\left\{\left|3,-1\right>,\,\left|3,1\right>\right\}$ and $\left|3,0\right>\rightarrow\left\{\left|2,-1\right>,\,\left|2,1\right>\right\}$ transitions, thereby increasing the energy of $\left|a\right>\equiv\left|2,0\right>$ and decreasing the energy of $\left|b\right>\equiv\left|3,0\right>$. From the Clebsch-Gordan coefficients for these transitions, we determine $c_\perp \simeq (0.357-0.092 \sin 2\epsilon)\,\mathrm{Hz/(\mu T)^2}$ at $f_\mathrm{s}=f_0$. A $B_\|$ component near $f_0$ is red detuned from the $\left|3,0\right>\rightarrow\left|2,0\right>$ transition and also increases the $\left|a\right>\leftrightarrow\left|b\right>$ transition frequency: $c_\| \simeq 0.468\,\mathrm{Hz/(\mu T)^2}$ at $f_\mathrm{s}=f_0$. Both $c_\perp$ and $c_\|$ are positive, of similar magnitude for all $\epsilon$ and depend only weakly on $f_\mathrm{s}$ around $f_0$ within the useful range of motional sideband frequencies, so this choice of $\left|a\right>$ and $\left|b\right>$ can be used to minimize both $B_\perp$ and $B_\|$.

Another useful choice is $\left|a\right>\equiv\left|2,0\right>$ and $\left|b\right>\equiv\left|3,1\right>$. Here, the dominant ac Zeeman shift results from $B_\|$ coupling to the $\left|3,1\right>\rightarrow\left|2,1\right>$ transition: $c_\| \simeq -49.08/ ((f_\mathrm{s}-f_0)/\mathrm{MHz})\,\mathrm{Hz/(\mu T)^2}$, and $c_\perp \simeq (0.199+0.220\sin 2\epsilon)\,\mathrm{Hz/(\mu T)^2}$ at $f_s=f_0$. This transition can be used to minimize $B_\|$ even further due to its large $c_\|$ value, by use of currents in electrodes MW2 and MW3 (those electrodes contribute dominantly to $B_\|$ due to their geometry). We produce the data of Fig.~\ref{fig:gradmap}c using this transition and an experimental sequence presented in Fig.~\ref{fig:gradmap}b. With $\bold{B_{\text{MW}}}\equiv0$ at position $\bold{\delta r} = \{\delta x, \delta y, \delta z\} = \{0, 0, 0\} $ in the coordinate system of Fig.~\ref{fig:chip}c [not the frame of Eq.~(\ref{eq:oscbfieldwrB})] we expect to first order
\begin{align}
\label{eq:oscbfield}
\nonumber
\bold{B_{\text{MW}}}=&B^\prime \cos(2\pi f_\mathrm{s} t)
\\
&\times \left(
    \begin{array}{ccc}
      \cos(2 \alpha_{\text{MW}} ) & 0 & \sin(2 \alpha_{\text{MW}} ) \\
      0           & 0 & 0           \\
      \sin(2 \alpha_{\text{MW}} ) & 0 &  -\cos(2 \alpha_{\text{MW}} )
    \end{array}
  \right)
  \,
  \left(
    \begin{array}{c}
      \delta x\\
      \delta y\\
      \delta z
    \end{array}
  \right).
\end{align}
Here $B^\prime$ and $\alpha_{\text{MW}}$ characterize the strength and orientation of the quadrupole for small displacements $\delta_{r} < 1.5~\mu$m from the field null point. By fitting the ac Zeeman shift resulting from the projection $B_\|$ of $\bold{B_{\text{MW}}}$ onto $\bold{B_{\text{0}}}$ to the experimental data of Fig.~\ref{fig:gradmap}c, we extract $B' = 35.3(4)$~T/m and $\alpha_{\text{MW}} = -26.5(7)^{\circ}$. From this fit, we determine a small shift of the center of the magnetic quadrupole with respect to the trap center by  $\{ \delta x, \delta z \} \simeq \{-0.11,0.23\}~\mu$m, limited by the sensitivity of our nulling procedure. We detect phase accumulations over a time $T_{\text{p}} = 250~\mu$s; we cannot measure phase shifts smaller than $\pi/10$ during that time. This limits the ac Zeeman shift sensitivity to $\simeq 0.2$~kHz and leads to finite fields at the trap center. The experimental day-to-day variations for the nulled configuration are $\leq 0.016$~dB for the relative powers of MW2 and MW3 with respect to MW1, and $\leq 0.12^{\circ}$ for the relative phases of MW1 and MW3 with respect to MW2.

Numerical simulations of the microwave currents in the three electrodes by use of ANSYS HFSS software~\cite{Note1} provide a value of $B^{\prime}_{\text{sim,1A}} \simeq 44$~T/m for a current amplitude of $1.0$~A in conductor MW1, while the amplitudes and phases of currents in MW2 and MW3 are adjusted to give a nulled field configuration at the position of the trap rf null. We find the following current amplitude ratios $R$ and phase shifts $\Delta \Phi$ in MW2 and MW3 relative to the current in MW1 for the nulled configuration: $\{R_{\text MW2}, R_{\text MW3}\} \simeq \{0.82, 0.96\}$ and $\{ \Delta \Phi_{\text{MW2}}, \Delta \Phi_{\text{MW3}} \} \simeq \{161.6^{\circ},-14.6^{\circ}\}$. A comparison with the experimental current ratios is difficult because of slightly varying attenuations in the signal paths leading to the three microwave electrodes. However, from simulations, we extract an angle $\alpha^{\text{sim}}_{\text{MW}} \simeq - 25^{\circ}$, close to the value fitted to the experimental data (also Sec.~\ref{sec:MM}). A similar level of agreement between simulation and experiment is found in~\cite{allcock_microfabricated_2012}. In our experiments, current densities are simulated to be as high as $\simeq 2 \times 10^{10}~\text{A/m}^2$ near the edges of the conductor MW1, with a total current of $\simeq 0.8$~A in MW1, for the above stated experimental settings, producing a gradient of $35$~T/m. We estimate the corresponding power dissipation in all electrodes due to all currents from the bulk resistance of gold ($\simeq 2.2 \times 10^{-8}~\Omega \cdot$m) in the central region of the chip ($\pm~200~\mu$m along $y$) to be $\simeq 80$~mW.
\begin{figure}
  \centering
  \includegraphics[width=\columnwidth]{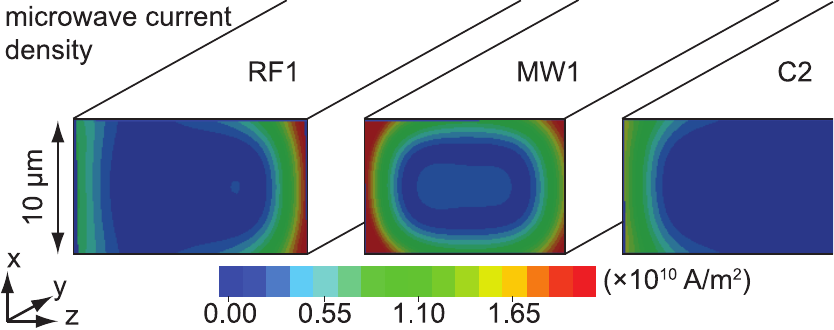}
  \caption{(Color) Cross-section view of the simulated microwave current density distribution in the trap electrodes when the magnetic field is nulled at the ion position. (only showing electrodes RF1, MW1, and C2). The densities shown here correspond to a gradient of $B^{\prime}_{\text{sim}} \simeq 35$~T/m. Currents in microwave electrodes induce currents in neighboring electrodes, here illustrated for RF1 and C2.} 
\label{fig:mwdensity}
\end{figure}
In Fig.~\ref{fig:mwdensity} the current density distribution in electrode MW1 and the induced-current densities in the neighboring electrodes RF1 and C2 are shown. The simulated magnetic and electric microwave fields for the nulled configuration are plotted in Fig.~\ref{fig:mwfields}. The influence of the microwave electric fields on the trapping potential is discussed in Appendix~\ref{sec:mwtrap}.
\begin{figure}
  \centering
  \includegraphics[width=\columnwidth]{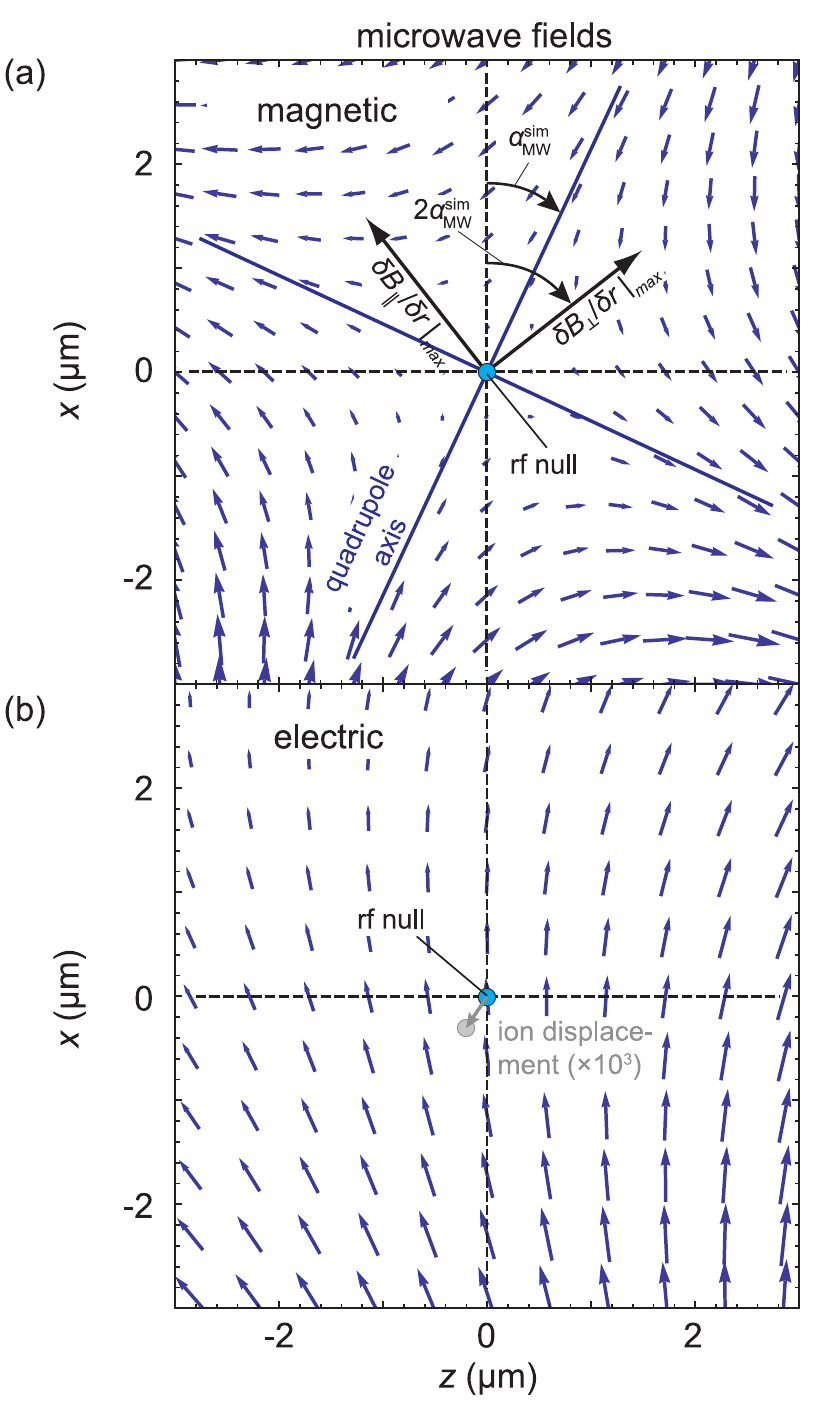}
  \caption{(Color online) Illustration of the simulated microwave fields in the radial plane. (a) Magnetic quadrupole field as a function of displacement from the trap center. (b) Electric field. Due to the pseudopotential from the electric field the ion(s) are displaced from the trap rf null as indicated in gray. The pseudopotential from the electric field displaces the ion(s) from the trap rf null (see Appendix~\ref{sec:mwtrap}). The estimated $\simeq 0.4$ nm displacement for typical trap parameters is shown in gray, scaled by a factor $10^3$ for visibility.} 
\label{fig:mwfields}
\end{figure}
\subsection{Microwave Approach for rf Micromotion Nulling}
\label{sec:MM}
In the presence of excess rf micromotion, the microwave magnetic near-field gradient can also drive micromotion-induced sidebands on carrier transitions~\cite{berkeland_minimization_1998}. This effect can be used to null micromotion along any direction in the radial $x$-$z$ plane (refer to the frame of Fig.~\ref{fig:chip}b). Therefore, it may have an advantage over certain optical schemes that are sensitive to micromotion only along the direction of a laser beam (or along the direction of the effective $k$-vector for stimulated-Raman transitions)~\cite{berkeland_minimization_1998}. For example, in a surface-electrode trap, the direction perpendicular to the surface can be difficult to probe with laser beams, because scattered light from the trap surface may interfere with detection and/or cause electrode charging.

The micromotion can be written as $(x_{\text{MM}} \bold{e}_{x} + y_{\text{MM}} \bold{e}_{y} + z_{\text{MM}} \bold{e}_{z}) \cos (2\pi f_\text{RF}t)$. In the presence of a microwave magnetic field gradient of frequency $f_{\text{s}}$ and in the coordinate frame of Fig.~\ref{fig:chip}c, the ion in its rest frame experiences fields with frequencies $f_\text{s} \pm f_\text{RF}$ and amplitude $\bold{\hat{B}}_{\text{ion}}$, cf. Eq.~(\ref{eq:oscbfield}):
\begin{align}
\label{ref:MMfield}
  \bold{\hat{B}}_{\text{ion}} = \frac{1}{2} B'
  \left(
    \begin{array}{ccc}
      \cos(2 \alpha_{\text{MW}}) & 0 & \sin(2 \alpha_{\text{MW}}) \\
      0 & 0 & 0\\
      \sin(2 \alpha_{\text{MW}}) & 0 &  -\cos(2 \alpha_{\text{MW}})
    \end{array}
  \right)
  \,
  \left(
    \begin{array}{c}
      x_{\text{MM}}\\
      0\\
      z_{\text{MM}}
    \end{array}
  \right).
\end{align}
For $f_\mathrm{s}=f_\mathrm{HFS} \pm f_\mathrm{RF}$, where $f_\mathrm{HFS}$ is the frequency of a transition between states in the hyperfine manifold, the ion experiences an oscillating field at frequency $f_\mathrm{HFS}$. The resulting Rabi rate is proportional to $|\bold{\hat{B}}_{\text{ion}}|$ and therefore to $r_{\text{MM}}$, the micromotion amplitude. To minimize the excess micromotion, we move the ion in the radial plane (along the shim directions illustrated in Fig.~\ref{fig:rffield}) to a position that minimizes the Rabi rate of the micromotion sidebands. The oscillating field $ \bold{\hat{B}}_{\text{ion}}$ has projections $B_\|$ and $B_\perp$ on $ \bold{B}_{\text{0}}$. Therefore, we can sense micromotion along any direction in the $xz$ plane, since it is always detectable by either a $\Delta m_{F}=0$ or $\Delta m_{F}=\pm 1$ transition (or both). In the experiment, we use the $\left|2,0\right>\leftrightarrow\left|3,1\right>$ and $\left|3,1\right>\leftrightarrow\left|2,1\right>$ transitions, with a magnetic-dipole matrix element $\simeq 0.414~\mu_{\text{B}}$ and $\simeq 1.001~\mu_{\text{B}}$, respectively, where $\mu_{\text{B}}$ denotes the Bohr magneton. 

To estimate the sensitivity of this nulling method, we assume $B'=35$~T/m and a transition matrix element for $f_\mathrm{HFS}$ of $\mu_\mathrm{B}$. To further simplify the estimate, the rotation angles of the microwave-magnetic quadrupole ($\alpha_{\text{MW}} \simeq -30^{\circ}$, Fig.~\ref{fig:mwfields}a) and rf-electric quadrupole ($\alpha_{\text{RF}} \simeq -20^{\circ}$, Fig.~\ref{fig:rffield}) are assumed to be oriented such that in the rest frame of the ion, we achieve a maximum oscillating field component that drives the transition ($\alpha_{\text{MW}} = \alpha_{\text{RF}}/2$). For the given misalignment, this approximation is good to within $15~\%$ for the estimated sensitivity. Note that in the experiments both components of the microwave-field gradient can be used, and the misalignment does not limit the sensitivity of this method. The Rabi rate for the micromotion sideband-induced carrier transition is then given by~\cite{ospelkaus_trapped-ion_2008}:
\begin{align}
  \Omega_{\text{MM}} &= \frac 1 2 B' r_{\text{MM}} \frac{\mu_{\mathrm{B}}}{2\hbar}.
\end{align}
For our parameters, and assuming a $\pi/10$ rotation detection sensitivity on the micromotion sideband during a drive time $T_{\text{p}}=2~$ms, a micromotion amplitude of $r_{\text{MM}}=0.1$~nm can be detected.

\begin{figure}
  \centering
  \includegraphics[width=\columnwidth]{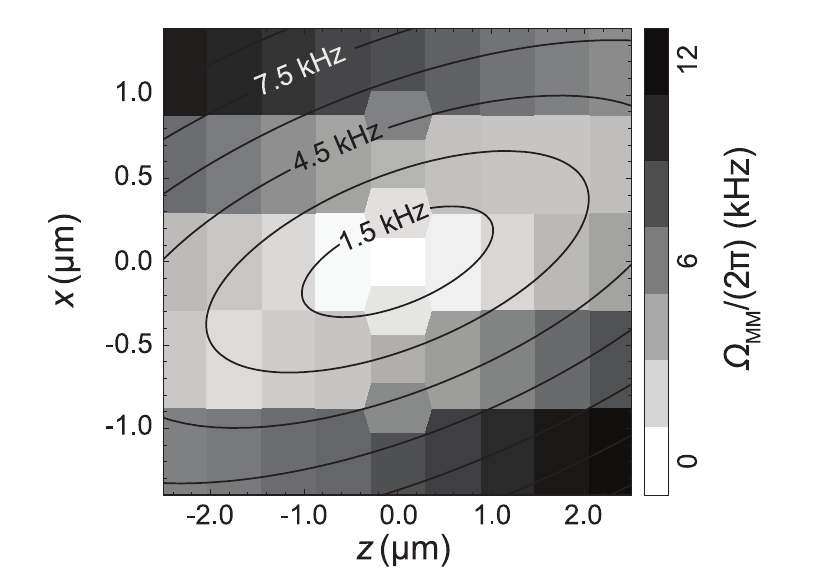}
  \caption{Rabi rate on the red micromotion sideband of the $\left|2,0\right>\leftrightarrow\left|3,1\right>$ transition as a function of ion position. The isolines result from a model fit to the data, from which  $B'=35.1(4)\,\mathrm{T/m}$ and $\alpha_{\text{MW}} = -31.1(2)^\circ$ are extracted from Eq.~(\ref{ref:MMfield}).
  }
\label{fig:mmmap}
\end{figure}

In addition to detecting micromotion, we use this method to probe characteristics of the microwave-magnetic and rf-electric field by measuring the micromotion sideband Rabi rate as a function of position. Figure~\ref{fig:mmmap} shows the Rabi rate on the red micromotion sideband of the $\left|2,0\right>\leftrightarrow\left|3,1\right>$ transition versus ion position.
Using these data together with the calculated shape of the rf quadrupole, we extract the parameters $B'$ and $\alpha_{\text{MW}}$ of the microwave quadrupole field, finding $B'=35.1(4)\,\mathrm{T/m}$ and $\alpha_{\text{MW}} = -31.1(2)^\circ$. This map was taken with a different realization of the trap described in Sec.~\ref{sec:OscB} but of the same design, so imperfections in construction and small variations in the electrode impedance may have affected the measurement. Differences in impedance affect the strength of the gradient, so the agreement in $B'$ values is probably fortuitous.  However, we find again a value for $\alpha_{\text{MW}}$ that is close to the simulated value $\alpha^{\text{sim}}_{\text{MW}} \simeq -25^\circ$ (Sec.~\ref{sec:OscB}).

\subsection{Control of the orientation of the radial motional modes}
\label{sec:radialMode}
\begin{figure}
  \centering
  \includegraphics[width=\columnwidth]{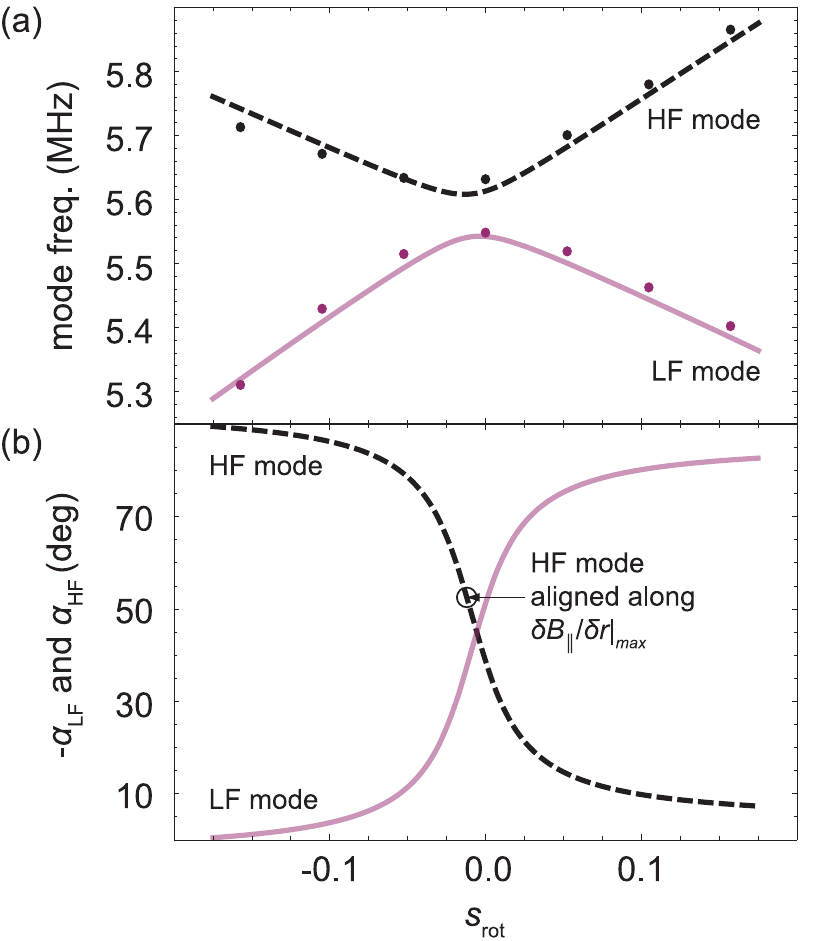}
  \caption{Control over the radial mode orientation. (a) Comparison of simulation (lines) with experimental data of radial-mode frequencies (error bars are too small to be visible) as a function of the scaling of the control field. (b) Mode angle $\alpha_{\text{LF}}$ extracted from the simulation. At an angle $\alpha_{\text{LF}} = 2 \alpha_{\text{MW}} \simeq-53^{\circ}$ the high-frequency mode is aligned along  $\delta B_\|/\delta r  |_{max}$ and couples maximally to the microwave-field gradient.
  }
\label{fig:radialcontrol}
\end{figure}

After determining the characteristics of the microwave quadrupole, we align the orientation of the radial motional modes to the microwave field gradient to optimize sideband Rabi rates. Since the qubit is controlled with a $\Delta m_{\text{F}} = 0$ transition, its motional sidebands are driven by $\delta B_\|/\delta r$ (Fig.~\ref{fig:mwfields}a). To overlap the HF-mode vector with $\delta B_\|/\delta r |_{max}$, we use an additional control potential $\phi_{\text{rot}}$ with potentials $V^{\text{rot}}_{\text{C1-C6}} = \{-1.694 ,2.298 ,-1.694, 0.311, 0.501, 0.311\}$~V applied to the control electrodes, which leads to a mode orientation $\alpha_{\text{LF}} \simeq 5^{\circ}$ in the absence of any additional control potential. In the experiments, the total potential $\phi_{\text{total}} =\phi_{\text{trap}} +\phi_{\text{stray}} + s_{\text{rot}} \phi_{\text{rot}}$ breaks the symmetry of the rf pseudopotential, and sets the orientation of the radial modes. The value of the scaling coefficient $s_{\text{rot}}$ is used to adjust the orientation of the radial modes. For the given total potential $\phi_{\text{total}}$ including the stray potential (Appendix~\ref{sec:trapPerform}), negative $s_{\text{rot}}$ rotates the LF-mode vector toward $\alpha_{\text{LF}} \rightarrow 0^{\circ}$ and for positive values it rotates toward $\alpha_{\text{LF}} \rightarrow -90^{\circ}$. Furthermore, $\phi_{\text{rot}}$ is designed such that the ion position remains unchanged while all trap frequencies change as a function of the scaling. A comparison of the simulated frequency changes with the actual experimental data can be used to determine and adjust the orientation of the radial modes. In Fig.~\ref{fig:radialcontrol}a such a comparison between simulated and measured mode frequencies shows good agreement. 
Figure~\ref{fig:radialcontrol}b illustrates the corresponding mode-angle values extracted from the simulation. Note that for the simulation we had to include $\phi_{\text{stray}}$ to find agreement with the experimental data, details are given in Appendix~\ref{sec:trapPerform}. When choosing $\alpha_{\text{LF}} = 2 \alpha_{\text{MW}} \simeq-53^{\circ}$, we observe the maximal coupling of the microwave-field gradient to the HF mode, while the LF mode couples very weakly to the microwave field. We find a corresponding Rabi rate for driving the HF sideband transition of the motional ground state of $\simeq 2\pi \times 2.0$~kHz, with the microwave-field gradient adjusted to $35$~T/m.

\section{Present limitations for two-qubit gates}
\label{sec:limits}
Internal-to-motional state coupling as well as an entangling two-qubit gate based on the simultaneous application of detuned red and blue sideband drives~\cite{soerensen_quantum_1999} was presented in~\cite{ospelkaus_microwave_2011}. Using this gate, entangled states with a fidelity of approximately $0.76$ were produced. In these experiments we used an out-of-phase radial mode of two ions, because it is less sensitive to electric-field noise than the corresponding center-of-mass mode. The out-of-phase motional heating was measured to be $0.2-0.5$~quanta/ms, while the center-of-mass motional heating rate was approximately $5$~quanta/ms (for radial mode frequencies around $7$~MHz). Heating corresponding to an increase of a single quantum of motion on the mode used for an entangling-gate operation effectively decoheres the gate operation~\cite{soerensen_entanglement_2000}. Since in~\cite{ospelkaus_microwave_2011} the entangling gate durations were approximately $250~\mu$s, even the out-of-phase mode heating can cause substantial gate errors. To investigate this effect for our experimental conditions, we numerically solve the master equation given in~\cite{soerensen_entanglement_2000}. We determine that this effect alone limits the fidelity to approximately $0.8$. Furthermore, we observed drifts of the motional sideband frequencies in the experiments, which may be caused by mode frequency drifts and/or varying ac Zeeman shifts from drifts of the residual magnetic and electric microwave fields. We estimate the frequency variation during the gate pulse to be less than $0.5$~kHz from these effects, while a typical detuning for the gate is around 4.5~kHz. This systematic effect can contribute an additional $5~\%$ loss of fidelity.

\section{Conclusion}
\label{sec:concl}
In summary, we have given details of the surface-electrode trap and microwave electronics used for the microwave near-field quantum control experiments described in~\cite{ospelkaus_microwave_2011}. We have described the experimental procedures used to control, shape and calibrate the microwave fields that implement motional-sideband transitions using the near-field gradients. We have introduced a novel technique for rf micromotion detection. Furthermore, we presented and experimentally verified a method to rotate the radial mode axes. Finally, we investigated limitations in fidelity of the two-qubit gates in~\cite{ospelkaus_microwave_2011} through simulations, and examined the impact of leading error sources. We determined that the motional heating during the gate pulse is the dominant source of infidelity. A significant reduction (by a factor of 100 or more) of the heating rate should lead to entangled state fidelities higher than $0.9$. A reduction of ambient heating may be achieved by surface treatments of the electrode structure as described in~\cite{allcock_reduction_2011,hite_100-fold_2012} and/or cryogenic cooling of the electrodes~\cite{deslauriers_scaling_2006, labaziewicz_suppression_2008, labaziewicz_temperature_2008}. In addition, higher microwave currents can increase the gate speed and thereby can reduce the effect of motional heating. The corresponding increased heat load may be mitigated by reducing electrode resistance through low temperature operation.

\begin{acknowledgments}
We thank M.~J.~Biercuk, J.~J.~Bollinger, and A.~P.~VanDevender for experimental assistance, and D.~H. Slichter and R. J\"ordens for helpful comments on the manuscript. We also thank D. Hanneke and J.~P.~Home for useful discussions and P. Treutlein for advice on microfabrication techniques. We received valuable assistance with microwave current simulations from D.T.C. Allcock and J. Sch\"obel. This work was supported by IARPA, ARO contract \#EAO139840, ONR, DARPA, Sandia National Laboratories and the NIST Quantum Information Program. Contribution of NIST, not subject to U.~S.~copyright.
\end{acknowledgments}

\begin{appendix}

\section{Comparison of Trap Simulation with Measurements}
\label{sec:trapPerform}
 
To compare the trapping potential calculations from Sec.~\ref{sec:trap} to the experiment, we first determine the electric field strengths at the position of the ion to compensate the rf micromotion (Sec.~\ref{sec:procedure}). We then measure the mode frequencies, and find the orientation of the radial mode vectors (Sec.~\ref{sec:radialMode}), characterized by $\alpha_{\text{LF}}$ (Fig.~\ref{fig:trappot}). We find a stray field $\bold{E}_{\text{stray}} \simeq 720 \times \{ -0.54,0.10, -0.84\}$~V/m and a stray field curvature tensor, given by the Hessian matrix of $\phi_{\text{stray}}$ at the center of the trap,
\begin{align}
\label{ref:stray}
\nonumber
H_{\text{stray}} \simeq 
  \left(
    \begin{array}{ccc}
      1.96 & 0.44 & 0.39\\
      0.44 & -1.67 & -4.23\\
      0.39 & -4.23 & -0.29
    \end{array}
  \right)\times 10^{7}~\text{V/m}^2.
\end{align}
After including $\phi_{\text{stray}}$ into the potential simulations, we find reasonably good agreement (motional frequencies and mode angles agree to within $10~\%$ or better) between the simulations and the measurements (Sec.~\ref{sec:radialMode}) for several different settings of rf voltage and control potentials. The stray potential might be caused by work function variations on the electrode surface~\cite{camp_macroscopic_1991, harlander_trapped-ion_2010, wilson_fiber-coupled_2011, brama_heating_2012} (e.g., from varying Mg contamination) or stray charges localized in the central region of the trap chip. One way of modeling modeling the stray effects is a rectangular patch with size $\Delta y = 35~\mu$m and $\Delta z = 5~\mu$m located on the chip surface, whose center is displaced by $z = - 25~\mu$m, $y = + 3~\mu$m relative to the trap center and that is biased to $-1.15$~V (typical work function differences can exceed one volt). We observe temporal variations in the stray potential effects, perhaps due to uncontrolled charging of the electrodes which may be caused by laser light impinging on the electrodes and/or changing Mg coverage of the chip surface during loading of the trap. In particular, the applied compensation fields vary within a few tens of volts per meter on a day-to-day basis, and the axial-mode frequency can vary more than $200$~kHz over the course of a few weeks with many loading cycles. 

We estimate the deviation of ion displacements from simulated values due to stray-electric fields to be less than a few percent for typical displacements from the rf null of $\pm 1~\mu$m at radial-mode frequencies of $\simeq 7.0$~MHz. We do not include this effect in our analysis of the experiments described in Sec.~\ref{sec:OscB} and Sec.~\ref{sec:MM}. However, for the calibrations described in Sec.~\ref{sec:radialMode}, we take the stray-electric field into account because of its significant impact on the radial motional-mode vector directions.
\section{Influence of the microwave electric fields on the trapping potential}
\label{sec:mwtrap}
In addition to the oscillating magnetic field $\bold{B}_{\text{MW}}$, an electric field $\bold{E}_{\text{MW}}$ (at the microwave frequency) is also present (Fig~\ref{fig:mwfields}b). If the null points for $\bold{E}_{\text{MW}}$ and $\bold{B}_{\text{MW}}$ fields differ, the pseudopotential produced by $\bold{E}_{\text{MW}}$ can push the ions away from the main trap rf pseudopotential null. From numerical simulations, we determine that the ion position in a $7.0$~MHz trap is shifted by $\{ \delta x, \delta z \} \simeq \{ 0.3, -0.2 \}$~nm when $B' \simeq 35$~T/m (Fig.~\ref{fig:mwfields}b). This displacement is much smaller than the precision for positioning the ion with the shim fields (Sec.~\ref{sec:trap}) and therefore this shift cannot be resolved. 

Another systematic effect is caused by the curvature of the microwave pseudopotential. It modifies the mode frequency (based on simulations) by about $0.5$~kHz, depending on the mode direction. This can be a significant shift for the two-qubit gate. However, when calibrating the sideband frequencies this effect is included, since we probe the sidebands with the microwave drive, which inherently includes the microwave pseudopotential produced by the gate pulse. To experimentally check for this shift, we measure the mode frequencies by applying a oscillating potential (coherent excitation) to one of the control electrodes, while the ion is in the $\left |3,3 \right >$ state. We then detect the ion fluorescence as a function of excitation frequency: a decrease in the resulting fluorescence indicates that a mode of the ion's motion has been resonantly excited~\cite{wineland_laser-fluorescence_1983}. We compare the measured mode frequencies with and without the microwave fields while applying the excitation pulse. With the microwaves applied, we observe shifts of the radial-mode frequencies of $\simeq 3$~kHz. The observed shift is about seven times higher than what we expect from our estimate; this discrepancy is not understood. The mode frequency shift (for a $7.0$~MHz trap) for the ion in the $\left |3,3 \right >$ state due to the spatially varying ac Zeeman shift is estimated to be less than $\simeq 10$~mHz and can be neglected here.

We checked for motional excitations when simultaneously applying two tones near $f_{\text{0}} \pm f_{\text{radial}}$, as required for the two-qubit gate (Sec.~\ref{sec:limits}). For pulse lengths up to $1$~ms (four times longer than typical gate times) we observe no change in the motional state within an uncertainty of approximately $\pm 0.5$~quanta. However, when we apply the two tones at $f_{\text{0}} \pm f_{\text{radial}}/2$, we record coherent motional excitations corresponding to a few tens of quanta after $50~\mu$s. The origin of the observed excitation may be the gradients of the two microwave pseudopotentials (with difference frequency $f_{\text{radial}}$). We estimate this effect to be negligible for our experimental conditions. However, we cannot rule out that the excitation is caused directly by an electric field at frequency $f_{\text{radial}}$. This field may originate from mixing of the two microwave tones due to nonlinear behavior of one or more of the microwave elements.

\end{appendix}

\end{document}